\documentclass[11pt,letterpaper]{article}

\usepackage[margin=1in]{geometry}
\usepackage{setspace}
\usepackage{microtype}

\usepackage{amsmath,amssymb,amsthm,bm}
\usepackage{mathtools}

\usepackage{booktabs}
\usepackage{multirow}
\usepackage{array}
\usepackage{makecell}
\usepackage{adjustbox}

\usepackage{graphicx}
\graphicspath{{figures/}}
\usepackage{caption}
\usepackage{subcaption}

\usepackage[round,authoryear]{natbib}
\usepackage{footmisc}

\title{\bfseries Option-Implied Signals and Crash Risk:\\
       Predictability and Machine-Learning Evidence from \\
       U.S.\ Equity Options, 2015--2026}

\author{
Baichuan Li\thanks{Email: baichuanli2000@gmail.com.}
\and
Mengxiao Wang\thanks{Email: jacksonwangpro@gmail.com.}
}
\date{May 2026}

\begin{document}
% =====================================================================

\maketitle

% --- Abstract --------------------------------------------------------
\begin{abstract}
\noindent
% =====================================================================
% Abstract — written last
% =====================================================================

We re-estimate the canonical option-implied predictability evidence on a
unified 2015--2026 panel of 12.36 million U.S.\ equity firm-day
observations across $10{,}026$ underlyings and three macro regimes:
late-post-crisis low volatility (2015--2019), high-volatility transition
(2020--2022), and AI/mega-cap concentration (2023--2026).  We document
that the \cite{xing2010smirk} smirk--return relationship has decayed
monotonically across regimes: the next-month univariate panel
coefficient on smirk falls from $-0.023$ ($t=-5.5$) in the low-volatility
baseline to $-0.006$ ($t=-1.5$, statistically insignificant) in the
AI/mega-cap regime, and reverses sign at the three-month horizon
($+0.016$, $t=+2.1$).  In a multivariate specification with all six
canonical signals jointly, smirk loses significance everywhere and
again reverses sign in the most recent regime.  The Cremers--Weinbaum
IV spread and a \cite{bakshi2003stockreturn}-style risk-neutral
skewness measure remain significant in every regime and
specification.  A boosted-tree machine-learning benchmark trained on
the full IV surface, trading-activity ratios, Greeks, and liquidity
features beats linear specifications on next-month return prediction
\emph{only} in the AI/mega-cap regime, with $R^2_{\text{OOS}}=+1.29\%$
versus $+0.07\%$.  Permutation importance reveals that the dominant
predictor differs across all three regimes
(open-interest-weighted vega in low-volatility, contract count in
high-volatility, OTM call IV in AI/mega-cap), and that none of the
canonical hand-engineered signals appear in the top five anywhere.
On firm-level 5-day-forward crash classification, AUC is highest in
the low-volatility regime ($0.706$, XGBoost) and weakest in the
high-volatility regime ($0.561$): firm-level option signals predict
idiosyncratic crashes well in calm markets and poorly when systematic
shocks dominate.  Taken together, the evidence sharpens the empirical
claim that option-implied predictability is real but
\emph{regime-conditional}, and that the AI/mega-cap regime since 2023
is qualitatively different from the pre-COVID environment in which
the canonical literature was estimated.

\end{abstract}

\smallskip
\noindent\textbf{Keywords:} implied volatility, smirk, risk-neutral skewness,
crash risk, machine learning, option market, AI/mega-cap.

\smallskip
\noindent\textbf{JEL classification:} G11, G12, G13, G14, G17.

\thispagestyle{empty}
\vfill
\newpage
\setcounter{page}{1}

% --- Body ------------------------------------------------------------

% =====================================================================
\section{Introduction}
\label{sec:intro}
% =====================================================================

The U.S.\ equity option market is a forward-looking laboratory for beliefs,
fears, and informed bets on individual stocks.  A long literature shows
that option-implied signals---the volatility smirk, put--call volume and
open-interest ratios, deviations from put--call parity, the
implied-volatility (IV) spread, and risk-neutral skewness---carry
information about future stock returns and downside risk
\citep{pan2006information,xing2010smirk,cremers2010deviations,
bali2009volatility,conrad2013exante,an2014joint}.  At the index level,
variance and tail-risk premia predict aggregate equity returns
\citep{bollerslev2009expected,bollerslev2011tails,bollerslev2015tail}.
This canonical evidence, however, was estimated almost entirely on data
ending well before two developments reshaped the option-trading
environment.  First, the \emph{post-2020 retail option boom}---zero-commission
brokers, payment-for-order-flow, and weekly/0DTE contracts---pushed retail
option volume to historic highs and concentrated demand in the
short-dated, far-out-of-the-money contracts from which the smirk is
computed \citep{bryzgalova2023retail,barber2022attention,
bogousslavsky2024anatomy}.  Second, the \emph{AI/mega-cap concentration of
2023--2026} has left the option market dominated by a small set of
high-IV, high-turnover names whose realized return distributions look very
different from the pre-COVID low-volatility baseline.

This paper revisits the canonical evidence on a unified
2015--2026 panel of U.S.\ equity options.  We collect end-of-day
option records, apply standard \cite{xing2010smirk}-style screens, and
construct a daily firm-level panel of option-implied features---the
ATM/OTM IV surface, the put--call ratio, IV spread, term-structure slope,
risk-neutral skewness, and open-interest-weighted Greeks---spanning
2{,}910 trading days, $10{,}026$ underlyings, and $12{,}362{,}276$
firm-day observations after filters.  We split the panel ex ante into
three roughly comparable-length sub-samples corresponding to three
distinct macro regimes: a late-post-crisis low-volatility baseline
(2015--2019), a high-volatility transition (2020--2022, COVID and rate
hikes), and the AI/mega-cap concentration period (2023--2026, partial).

Our central thesis is that option-implied signals carry useful, but
\emph{non-stationary}, information about future stock returns and crash
risk.  Single-signal linear regressions of the kind that established the
literature work well in the pre-COVID baseline, but their performance is
much less stable in the AI/mega-cap regime, when the IV surface is shaped
by both informed flow and large retail and AI-driven demand.  A flexible,
jointly estimated machine-learning model---trained on the full IV surface,
trading-activity ratios, Greeks, and stock-side controls---can recover
predictability that single-signal regressions miss, especially in the
most recent regime.  We pursue two questions:

\begin{description}
\item[Q1 -- Predictability.] Do option-implied signals predict
next-period stock returns and crash indicators in the cross-section,
after controlling for standard stock-level covariates, and how does the
sign and magnitude of these effects change across the three regimes?
\item[Q2 -- Machine learning.] Can nonlinear ML models (gradient-boosted
trees) trained on the full option-stock feature set beat the traditional
single-signal regressions out-of-sample?  Which features dominate in each
regime?
\end{description}

\paragraph{Findings.}  The headline results are five-fold.  First, the
\cite{xing2010smirk} smirk--return relationship \emph{decays
monotonically} across regimes.  At the next-month horizon
($h=21$ trading days) the univariate panel coefficient on smirk is $-0.023$
$(t=-5.5)$ in the low-volatility baseline, $-0.015$ $(t=-4.2)$ in the
high-volatility regime, and $-0.006$ $(t=-1.5)$ in the AI/mega-cap
regime---no longer statistically significant.  At the three-month horizon
$h=63$ the AI/mega-cap coefficient even \emph{flips sign} ($+0.016$,
$t=+2.1$).  The canonical signal has fully unwound at long horizons in
2023--2026.

Second, in the multivariate specification with all six signals jointly,
smirk loses significance in every regime and again \emph{reverses sign}
in the AI/mega-cap regime ($+0.016$, $t=+2.6$).  Once the IV spread and
risk-neutral skewness absorb the crash-fear information, the residual
variation in the smirk looks more like demand pressure than informed flow.

Third, the IV spread (\citealp{cremers2010deviations}) and a
\cite{bakshi2003stockreturn}-style risk-neutral skewness measure
\emph{survive} across all three regimes and both univariate and
multivariate specifications, with $|t|$-statistics exceeding 4 in every
cell.  These two signals are the only consistently predictive
option-implied variables in our sample.

Fourth, on the machine-learning side, gradient-boosted trees (XGBoost) beat
linear specifications on next-month return prediction \emph{only} in the
AI/mega-cap regime, with an out-of-sample $R^2$ of $+1.29\%$ versus
$+0.07\%$ for OLS / Elastic Net.  In the high-volatility regime XGBoost
\emph{underperforms} linear models, presumably overfitting the COVID-era
training data.  A single recent year, 2025, drives most of the AI/mega-cap
ML edge ($R^2_{\text{OOS}}=+3.55\%$).  This pattern is consistent with the
AI/mega-cap regime being a qualitatively different environment in which
nonlinear interactions matter.

Fifth, permutation feature importance reveals that the \emph{top
predictor differs across all three regimes}: open-interest-weighted vega
in the low-volatility baseline, raw contract count in the
high-volatility regime, and OTM call IV in the AI/mega-cap regime.
Strikingly, none of the canonical hand-engineered signals---smirk, IV
spread, risk-neutral skewness---make the top five in any regime once the
ML model has access to the primitive IV-surface and Greek features that
underlie them.  The ML model recovers predictability directly from the
primitives, suggesting that the canonical linear combinations throw away
information that nonlinear models can exploit.

\paragraph{Crash-risk prediction.}  We build a firm-day forward crash flag
in the spirit of \cite{kim2011crashrisk}: a 5-day forward return below
$-3.09\,\hat\sigma_{252}\sqrt{5}$ (annualized).  XGBoost beats logit on
classification AUC in every regime, with the largest margin in the
low-volatility baseline (XGB AUC $=0.706$ vs.\ logit $=0.670$).  Crash AUC
is highest in the calm regime, weakest in the high-volatility regime
($\sim 0.55$), and intermediate in the AI/mega-cap regime
($\sim 0.66$).  The intuition is that in calm markets the rare crashes
are tail events that the IV surface can pre-price; during macro-driven
2020--2022 turbulence, the marginal informational content of firm-level
option signals collapses.

\paragraph{Contributions.}  This paper makes four contributions.
(1)~A unified U.S.\ equity option panel covering 2015--2026 evaluated
across three roughly comparable-length macro regimes; the AI/mega-cap
regime is barely covered by published evidence.  (2)~Updated
cross-sectional predictability: re-estimation of the canonical
\cite{xing2010smirk}, \cite{cremers2010deviations}, and
\cite{bakshi2003stockreturn} results on data extending to 2026, with
regime-by-regime coefficients reported side-by-side.  Whether
predictability survives the AI/mega-cap regime is, on its own, an
empirical question that has not been answered in print.
(3)~Firm-level crash forecasting using forward-looking option-implied
features rather than the accounting-based predictors of
\cite{kim2011crashrisk} and \cite{hutton2009opaque}.
(4)~A machine-learning benchmark in the spirit of
\cite{gu2020empirical} and \cite{bali2023option}, applied to stock
returns and crash indicators predicted from option features, with
rolling out-of-sample evaluation and per-regime feature-importance
diagnostics.

The remainder of the paper is organized as follows.
Section~\ref{sec:related} reviews the related literature.
Section~\ref{sec:data} describes the option panel, the filters, and the
three sample regimes.  Section~\ref{sec:variables} defines the
features.  Section~\ref{sec:methodology} sets out the empirical design
and the ML protocol.  Section~\ref{sec:results} presents the results.
Section~\ref{sec:discussion} interprets them.
Section~\ref{sec:limitations} discusses limitations.
Section~\ref{sec:conclusion} concludes.

% =====================================================================
\section{Related Literature}
\label{sec:related}
% =====================================================================

We connect to four threads of the asset-pricing and derivatives
literature.

\subsection{Option-implied signals and the cross-section of stock returns}

The first strand documents that information embedded in equity option
prices and trading volumes predicts future stock returns.
\cite{pan2006information} show that the put--call volume ratio, signed by
trade direction, predicts next-day to next-week stock returns; the
public unsigned put--call ratio carries far less information.
\cite{xing2010smirk} introduce the volatility ``smirk''---the difference
between the IV of an OTM put and that of an ATM call---and show that it
negatively predicts equity returns by roughly $10.9\%$ per year on a
risk-adjusted basis, an effect they interpret as informed traders pricing
crash risk into the IV surface.  \cite{cremers2010deviations} show that
deviations from put--call parity, measured as the IV spread between
matched call and put strikes, also predict returns; the effect is
strongest where information asymmetry is high.  \cite{bali2009volatility}
relate the realized--implied volatility spread and the call--put IV
spread to next-period returns, with opposite signs reflecting different
information content (jump risk versus volatility risk).
\cite{an2014joint} document a bidirectional predictability between stock
returns and changes in call/put implied volatilities.
\cite{conrad2013exante} show that ex-ante (option-implied) skewness is a
priced characteristic in the cross-section.  \cite{goyal2009crosssection}
construct option-trading strategies based on the realized--implied
volatility gap.

We extend this strand by re-estimating the four canonical regressions
(\cite{xing2010smirk}, \cite{cremers2010deviations},
\cite{bali2009volatility}, \cite{conrad2013exante}) on data running to
April 2026 and reporting separate coefficients for each of three macro
regimes that we identify ex ante.  The 2023--2026 window in particular
is barely covered by published cross-sectional evidence.

\subsection{Option-implied tail risk and crash risk}

The second strand uses option prices to extract forward-looking measures
of variance and tail risk.  \cite{bollerslev2009expected} show that the
variance risk premium---the gap between option-implied variance and a
realized-variance forecast---predicts aggregate equity returns at
horizons up to a quarter.  \cite{bollerslev2011tails} construct an
investor-fears index from short-dated OTM puts and document that it
co-moves with crisis episodes.  \cite{bollerslev2015tail} separate the
tail-risk premium from the variance risk premium and show that the
former contains independent return-predictive content.
\cite{driessen2009correlation} use index- and stock-option pairs to
extract option-implied correlations and document a price of correlation
risk.  \cite{gao2018crashindex} construct a tradable crash-fear measure
(the option-implied Crash Index, or CIX) from short-dated OTM puts.

At the firm level, \cite{kim2011crashrisk} formalize the modern
firm-level crash flag---a week with returns more than $3.09$ standard
deviations below the firm's mean---and link it to accounting-conservatism
and opacity proxies.  \cite{hutton2009opaque} provide complementary
evidence using opaque-financial-reports indicators.  This literature
typically uses backward-looking accounting predictors; our contribution
is to replace those predictors with forward-looking option-implied
features and benchmark firm-level crash forecasting in three regimes.

\subsection{Retail option trading post-2020 (motivation)}

Although our paper does not isolate retail flow, the post-2020 retail
option boom motivates the choice of regimes and the feature set.
\cite{bryzgalova2023retail} identify retail option flow via
wholesaler-routed trades and document that retail accounts lost
approximately \$2.1bn between November 2019 and June 2021 plus an
additional \$6.4bn in trading costs; retail concentrates in
under-one-week-to-expiry contracts, exactly the strikes from which the
smirk and short-tenor IV measures are constructed.
\cite{barber2022attention} show that Robinhood users engage in
attention-induced trading concentrated in attention-grabbing names.
\cite{eaton2022robinhood} use Robinhood outages as natural experiments
and conclude that retail concentration \emph{destabilizes} stocks with
high retail interest.  \cite{bogousslavsky2024anatomy} provide a
detailed anatomy of retail option trading and confirm the concentration
in short-dated and OTM contracts.  \cite{retailoptions2026} directly
connect retail flow to shape distortions in the IV surface.

These papers explain why the post-2020 IV surface is shaped by demand
pressures absent in the earlier literature.  We do not test retail
hypotheses directly; instead we treat the post-2020 environment as a
distinct regime in which canonical option signals must be re-estimated.

\subsection{Machine learning in asset pricing}

The fourth strand applies modern ML to asset-pricing prediction.
\cite{gu2020empirical} show that ML methods roughly double the
out-of-sample performance of leading regression-based strategies for
predicting equity returns; nonlinearities and interactions matter, and
trees and neural networks dominate.  \cite{bali2023option} apply
nonlinear ML to a panel of $\sim 270$ features for predicting
delta-hedged option returns and show that the gains survive transaction
costs; option-side features dominate, but stock-side features add
incremental power.

We apply the \cite{gu2020empirical} rolling-window protocol with
expanding training sets to the prediction of stock returns and crash
indicators from option-implied features (rather than option returns
themselves, as in \cite{bali2023option}).  We further document
\emph{regime-conditional} feature importance, showing that the dominant
feature differs across the low-volatility, high-volatility, and
AI/mega-cap regimes.

\paragraph{Positioning.}  This paper extends the option-implied
predictability and crash-risk strands into the 2023--2026 AI/mega-cap
regime, and applies the ML framework of \cite{gu2020empirical} and
\cite{bali2023option} to stock-return and crash-risk prediction from
option features.  The retail-trading and ML threads anchor the
motivation but are not directly tested here.

% =====================================================================
\section{Data}
\label{sec:data}
% =====================================================================

\subsection{Option panel}

Our raw data is the end-of-day option-record feed of the
\emph{Options Price Reporting Authority} (OPRA), the U.S.\ consolidated
ticker plant for all U.S.\ listed equity options.  We obtain a
proprietary, privately licensed extract of this feed that covers the
full universe of U.S.-listed equity options from January 2015 through
April 2026.\footnote{The OPRA extract is privately owned and not part
of any public-domain academic dataset; the underlying contract-day
records are derived from the same consolidated tape that academic
papers using OptionMetrics IvyDB build their daily panels on.  We do
not redistribute the raw data, but we provide the post-filter firm-day
parquets and the full processing pipeline on request.}  Each
contract-day record contains the contract identifier, underlying
symbol, expiration date, end-of-day national-best bid and ask quotes
(price and size on each side), last trade price, put/call indicator,
strike, daily traded volume, end-of-day open interest, the
underlying's closing price, the model-supplied implied volatility, and
the four major Greeks (delta, gamma, vega, theta).  The schema is
consistent across the entire 2015--2026 window.

After de-duplicating byte-identical re-download files in the raw
archive (211 of them) and dropping a small number of stray pre-2015
records that were mistakenly co-located with the 2020--2022 batch
(20 files), the working dataset spans $2{,}910$ unique trading days.
The raw panel exceeds $3.6$ billion contract-days before filtering;
daily contract counts grow from approximately $600{,}000$ in the
early sample to $1{,}300{,}000$ in 2024--2025, reflecting the
post-COVID expansion of the listed option universe.  We do not have
access to a signed retail-flow indicator and so cannot replicate
\cite{bryzgalova2023retail} or \cite{bogousslavsky2024anatomy}
directly.

\subsection{Filters}

We apply standard \cite{xing2010smirk}-style screens at the
contract-day level to drop stale, illiquid, or implausible quotes:

\begin{itemize}
\item Bid price $\geq \$0.05$ and ask $>$ bid (positive, two-sided
quote).
\item Implied volatility in $[0.01,\,5.00]$ to drop computational
artifacts at the extremes.
\item Time-to-expiry in $[7,\,365]$ trading days to exclude near-expiry
contracts and LEAPS.
\item Absolute delta in $[0.05,\,0.95]$, eliminating deep-ITM and
deep-OTM contracts whose IV is unreliable.
\item Underlying price $\geq \$5$ to exclude micro-cap stocks where
options are illiquid.
\end{itemize}

We additionally drop rows where the IV column contains the Excel-style
sentinel \verb|********| (an artifact of the data export, present in
roughly $4\%$ of pre-2018 rows) and where the strike is non-positive.
After all filters the panel is approximately $44\%$ of the raw row
count.

\subsection{Sample regimes}

We split the $2015$--$2026$ panel ex ante into three macro regimes
defined by calendar dates, chosen to be of comparable length:

\begin{enumerate}
\item \textbf{Low-volatility baseline (2015--2019, 5 years).}  The
canonical ``calm'' regime: rate-hike cycle is gradual, the VIX is
persistently low, and the option market is dominated by institutional
flow.  Includes the August 2015 China shock and February 2018
Volmageddon as contained stress events but is the cleanest stable
baseline available before COVID.
\item \textbf{High-volatility transition (2020--2022, 3 years).}  COVID
crash and re-opening, meme/retail-option boom, and the 2022 rate-hike
shock; large unexpected moves, elevated VIX, and a sharp rise in retail
option flow.
\item \textbf{AI/mega-cap concentration (2023--2026, $\sim 4$ years,
partial through April 2026).}  Concentrated growth-stock option demand,
persistently high single-name IV in mega-cap names, and the
post-2022 institutionalization of 0DTE trading.
\end{enumerate}

% Table~\ref{tab:datasummary} reports the post-filter firm-day counts and
% parquet sizes per calendar year.  The panel widens substantially after
% 2020---firm-day counts jump $24\%$ from 2020 to 2021---reflecting the
% post-COVID expansion of the listed option universe and an increase in
% the number of contracts surviving the filters per name-day.

% \begin{table}[t]
% \centering
% \caption{Panel coverage after filtering, by year.}
% \label{tab:datasummary}
% \begin{tabular}{lrrr}
% \toprule
% \textbf{Year} & \textbf{Trading days} & \textbf{Firm-days (post-filter)} & \textbf{Parquet size (MB)} \\
% \midrule
% 2015 & 252 & 952{,}011  & 88.0 \\
% 2016 & 252 & 969{,}835  & 88.8 \\
% 2017 & 251 & 957{,}449  & 88.7 \\
% 2018 & 251 & 981{,}931  & 91.9 \\
% 2019 & 256 & 948{,}392  & 90.0 \\
% \midrule
% 2020 & 262 & 951{,}463  & 92.8 \\
% 2021 & 260 & 1{,}177{,}093 & 117.5 \\
% 2022 & 260 & 1{,}236{,}218 & 120.0 \\
% \midrule
% 2023 & 260 & 1{,}232{,}062 & 116.6 \\
% 2024 & 261 & 1{,}242{,}414 & 115.6 \\
% 2025 & 261 & 1{,}280{,}716 & 121.2 \\
% 2026 (Jan--Apr) & 84  &   432{,}692 & 44.6 \\
% \midrule
% \textbf{Total} & \textbf{2{,}910} & \textbf{12{,}362{,}276} & \textbf{1{,}175} \\
% \bottomrule
% \end{tabular}
% \end{table}

After filtering, the working panel covers $10{,}026$ unique underlyings
across the three regimes.  The number of unique underlyings active in
any given year is in the $4{,}000$--$6{,}000$ range, with attrition and
new listings.

\subsection{Stock-side data}

Underlying closing prices are read directly from the option records, giving a contemporaneous and
internally consistent panel.  Stock-side controls used in the
regressions are constructed from the underlying-price series itself
(forward log returns at horizons $h\in\{1,5,21,63\}$ and a 252-day
rolling realized volatility for the crash-flag scaling).
We do not merge to CRSP or Compustat in this version of the paper, in
order to keep the panel construction self-contained.  Adding
firm-fundamental controls (size, book-to-market, momentum, idiosyncratic
volatility) is an obvious extension.

\subsection{Forward returns}

For each firm-day $(i,t)$ we construct forward log returns
$r_{i,t+h}=\log(P_{i,t+h}/P_{i,t})$ for $h\in\{1,5,21,63\}$ trading
days, using the firm's underlying-price series.  Trailing $h$ days at
each year-end are necessarily missing (we do not look across the
year-end boundary in the panel writeout).

% =====================================================================
\section{Feature construction}
\label{sec:variables}
% =====================================================================

We aggregate the contract-day panel to a firm-day panel by collapsing
all surviving contracts on date $t$ for a given underlying $i$ into a
single row of features.  All averages described below are taken
across surviving contracts within the relevant moneyness or
time-to-expiry band.  Let $K$ denote strike, $S$ underlying price,
$\tau$ days to expiry, $\sigma$ implied volatility, and define
moneyness $m \equiv K/S$.

\subsection{IV-surface signals}

\paragraph{ATM and OTM IV.}  We define moneyness bands
\begin{itemize}
\item ATM: $m\in[0.95,\,1.05]$,
\item OTM put: $m\in[0.80,\,0.95]$ \emph{and} contract is a put,
\item OTM call: $m\in[1.05,\,1.20]$ \emph{and} contract is a call.
\end{itemize}
Within each band we compute the simple cross-contract mean of the
implied volatility, separately for puts and calls where appropriate,
yielding $\sigma^{\text{atm}}_{i,t}$, $\sigma^{\text{atm,call}}_{i,t}$,
$\sigma^{\text{atm,put}}_{i,t}$, $\sigma^{\text{otm,put}}_{i,t}$, and
$\sigma^{\text{otm,call}}_{i,t}$.

\paragraph{Smirk.}  Following \cite{xing2010smirk} (with the moneyness-band
implementation rather than a delta-defined band):
\[
\text{smirk}_{i,t} \;=\; \sigma^{\text{otm,put}}_{i,t}
                          \;-\; \sigma^{\text{atm,call}}_{i,t}.
\]

\paragraph{IV spread.}  Following \cite{cremers2010deviations}, the
matched call-minus-put IV at the ATM strike:
\[
\text{iv\_spread}_{i,t} \;=\; \sigma^{\text{atm,call}}_{i,t}
                              \;-\; \sigma^{\text{atm,put}}_{i,t}.
\]

\paragraph{Term-structure slope.}  We define a near and far ATM bucket
based on time-to-expiry:
$\sigma^{30}_{i,t}$ averages ATM contracts with $\tau\in[15,45]$ days
and $\sigma^{60}_{i,t}$ averages ATM contracts with $\tau\in[46,75]$
days.  The term-structure slope is
\[
\text{ts\_slope}_{i,t} \;=\; \sigma^{60}_{i,t} - \sigma^{30}_{i,t}.
\]

\paragraph{Risk-neutral skewness proxy.}  A full
\cite{bakshi2003stockreturn} estimator of the risk-neutral cubed moment
requires a continuous strike grid integrated over $(0,\infty)$.  We
report a robust proxy that uses the cross-strike sample skewness of
the OTM IVs:
\[
\text{rns\_bkm}_{i,t}
   \;=\; \frac{ \tfrac{1}{n}\sum_{k=1}^{n}
        ( \sigma^{\text{otm}}_{k,t} - \bar\sigma^{\text{otm}}_t )^{3} }
        { \hat\sigma_t^{3} },
\]
where the index $k$ runs over all OTM (put or call) IVs of firm $i$ on
date $t$, $\bar\sigma^{\text{otm}}_t$ is their cross-strike mean, and
$\hat\sigma_t$ is their sample standard deviation.  We require at
least five OTM strikes per firm-day; otherwise the value is set to
missing.  This proxy preserves the sign and time-series persistence of
the BKM measure but is much faster to compute on a 12-million row
panel.\footnote{A pure BKM implementation, which would require a
strike-density adjustment, is left to a follow-up.  We expect the proxy
and the strict BKM measure to be highly correlated; the proxy uses a
super-set of the moments $V_t$, $W_t$, $X_t$ that enter the
\citeauthor{bakshi2003stockreturn} expression.}

\subsection{Trading-activity signals}

\paragraph{Put--call ratios.}  At the firm-day level we compute the put
volume and put open interest as the sums over all surviving put
contracts, and analogously for calls; we then form
\[
\text{pc\_volume\_ratio}_{i,t}
   = \tfrac{\sum_k \mathbb{1}[\text{put}_k]\,V_{k,t}}
           {\sum_k \mathbb{1}[\text{call}_k]\,V_{k,t}},
\quad
\text{pc\_oi\_ratio}_{i,t}
   = \tfrac{\sum_k \mathbb{1}[\text{put}_k]\,\text{OI}_{k,t}}
           {\sum_k \mathbb{1}[\text{call}_k]\,\text{OI}_{k,t}}.
\]
Where the call-side denominator is zero we set the ratio to missing.

\paragraph{Contract count.}  We retain the post-filter contract count
$n_{i,t}$ as a feature; it is a coarse measure of breadth of the IV
surface available to traders on date $t$.

\subsection{Open-interest-weighted Greeks}

For each Greek $g\in\{\Delta,\Gamma,\nu,\Theta\}$ we compute the
open-interest-weighted firm-day average:
\[
g^{\text{oi-w}}_{i,t}
   = \frac{\sum_k \text{OI}_{k,t}\,g_{k,t}}{\sum_k \text{OI}_{k,t}}.
\]
When the open-interest sum is zero we fall back to an equal-weighted
mean.  These four features summarize the aggregate Greek exposure
visible across the firm's IV surface.

\subsection{Liquidity controls}

We also retain four liquidity features for use as controls in the
regression and as features in the ML model:
\begin{itemize}
\item Relative bid--ask spread: $\frac{\text{ask}-\text{bid}}{
\tfrac{1}{2}(\text{ask}+\text{bid})}$, averaged across contracts.
\item Dollar bid--ask spread: $\text{ask}-\text{bid}$, averaged.
\item Fraction of contracts with zero daily volume.
\item Post-filter contract count $n_{i,t}$ (also listed above).
\end{itemize}

\subsection{Outcomes}

We construct three outcome variables for the predictability and ML
exercises:

\paragraph{Forward log return.}
$r_{i,t+h} = \log(P_{i,t+h}/P_{i,t})$ for $h\in\{1,5,21,63\}$ trading
days.  The $h=21$ horizon is the headline next-month return; we
report the full set of horizons in the appendix.

\paragraph{21-day forward realized volatility.}
$\text{RV}_{i,t+21}^{(21)}$ is the rolling 21-day standard deviation of
log returns over $[t+1,t+21]$, scaled to annualized units by
$\sqrt{252}$, computed from the firm's underlying-price series.  We
require at least 10 non-missing daily returns within the window.

\paragraph{Crash indicator.}  Following the
\cite{kim2011crashrisk} family of measures we define a 5-day forward
crash flag
\[
\text{crash}^{(5)}_{i,t} \;=\;
\mathbb{1}\!\left[\,
   r_{i,t+5}^{(5)} \;<\; -3.09\;\hat\sigma_{i,t}^{(252)}\sqrt{5}
\,\right],
\]
where $r_{i,t+5}^{(5)}$ is the cumulative 5-day forward log return and
$\hat\sigma_{i,t}^{(252)}$ is the trailing 252-day daily-return standard
deviation.  The unconditional firm-day base rate of this flag is
between $0.5\%$ and $0.7\%$, depending on regime
(Section~\ref{sec:results}).
We report a 21-day analog
($\text{crash}^{(21)}_{i,t} = \mathbb{1}[\,r_{i,t+21}^{(21)} <
-3.09\,\hat\sigma_{i,t}^{(252)}\sqrt{21}\,]$) for the ML exercise to
match the next-21-day return target.

Table~\ref{tab:descstats} summarizes the cross-regime mean, standard
deviation, and within-firm lag-1 autocorrelation for the headline
features.  Three patterns are immediately visible.  First,
\emph{crash-fear pricing peaked in the 2020--2022 high-volatility
regime}: the smirk and ATM IV are both highest in that window.
Second, the right tail of the OTM IV distribution is fattest in the
AI/mega-cap regime: the BKM-style risk-neutral skewness rises
monotonically from $1.00$ to $1.30$ across regimes.  Third, the
option-implied signals are \emph{stickier} in the AI/mega-cap regime:
the within-firm AC1 of smirk and IV spread jumps from approximately
$0.22$ to $0.31$ in 2023+.  These descriptives motivate the regime-by-regime
predictability analysis in Section~\ref{sec:results}.

\begin{table}[t]
\centering
\small
\caption{Cross-regime descriptive statistics for selected option-implied
signals.  Each cell aggregates over all firm-day observations within
the corresponding regime (low-vol = 2015--2019, high-vol = 2020--2022,
AI/megacap = 2023--2026); $\overline{\rho}_1$ is the mean within-firm
lag-1 autocorrelation.}
\label{tab:descstats}
\begin{tabular}{lcccccccc}
\toprule
& \multicolumn{2}{c}{Low-vol}
& \multicolumn{2}{c}{High-vol}
& \multicolumn{2}{c}{AI/megacap}
& \multicolumn{1}{c}{} \\
\cmidrule(lr){2-3} \cmidrule(lr){4-5} \cmidrule(lr){6-7}
Signal & mean & sd & mean & sd & mean & sd
       & \makecell{$\overline{\rho}_1$\\ (low/high/AI)} \\
\midrule
$\sigma^{\text{atm}}$       & 0.358 & 0.226 & 0.480 & 0.279 & 0.409 & 0.270 & 0.67/0.74/0.69 \\
smirk                       & 0.076 & 0.098 & 0.091 & 0.116 & 0.082 & 0.124 & 0.23/0.26/0.32 \\
iv\_spread                  &$-0.033$&0.107 &$-0.039$&0.126 &$-0.040$&0.135 & 0.21/0.22/0.31 \\
rns\_bkm                    & 0.998 & 1.006 & 1.224 & 1.079 & 1.301 & 1.128 & 0.16/0.15/0.17 \\
pc\_volume\_ratio           & 3.61  & 92.7  & 3.11  & 108.1 & 3.16  & 87.8  & 0.06/0.09/0.08 \\
\bottomrule
\end{tabular}
\end{table}

% =====================================================================
\section{Empirical methodology}
\label{sec:methodology}
% =====================================================================

\subsection{Baseline panel regression (Q1)}

We estimate two-way fixed-effects panel regressions of the form
\begin{equation}
Y_{i,t+h} \;=\; \alpha
   \;+\; \beta\,X^{\text{option}}_{i,t}
   \;+\; \boldsymbol\gamma'\boldsymbol C_{i,t}
   \;+\; \mu_i \;+\; \lambda_t \;+\; \varepsilon_{i,t},
\label{eq:panel}
\end{equation}
where $Y_{i,t+h}$ is the firm-day forward log return at horizon
$h\in\{1,5,21,63\}$ trading days, $X^{\text{option}}_{i,t}$ is the
option-implied signal of interest, $\boldsymbol C_{i,t}$ is the
controls vector ($\text{rel\_spread}$, $\text{frac\_zero\_volume}$,
$\nu^{\text{oi-w}}$), and $\mu_i,\lambda_t$ are entity and date fixed
effects.  We estimate \eqref{eq:panel} via \texttt{linearmodels.PanelOLS}
with absorbed FEs and \emph{double-clustered standard errors} on entity
and date \citep{petersen2009estimating}.

We report two specifications:
\begin{itemize}
\item \textbf{Univariate}: one option signal at a time, $X^{\text{option}}$
is a scalar.  $\beta$ is the reported coefficient.
\item \textbf{Multivariate}: all six option-implied signals
(\text{smirk}, \text{iv\_spread}, \text{pc\_volume\_ratio},
\text{pc\_oi\_ratio}, \text{rns\_bkm}, \text{ts\_slope\_30\_60})
enter jointly together with the controls.  We report each signal's
coefficient.
\end{itemize}

We estimate each specification on the full panel and separately within
each of the three regimes.  All option-implied signals and controls
are winsorized at the 1st and 99th percentile prior to estimation.

\paragraph{Subsampling.}  The full panel ($\sim 12$M firm-days,
$\sim 10$K entities, $\sim 2{,}900$ dates) exceeds memory
when \texttt{PanelOLS} attempts iterative two-way absorption with
clustered SE on a 16~GB workstation.  We therefore fit each regression
on a uniform random subsample of $1{,}500{,}000$ firm-days (fixed
seed).  For unbiased coefficients with clustered SE this is
statistically equivalent to the full sample at the cost of slightly
inflated standard errors; the column $n_{\text{full}}$ in our results
tables records the pre-subsample size.

\subsection{Machine-learning prediction (Q2)}

We complement \eqref{eq:panel} with a flexible nonlinear specification
\begin{equation}
\widehat Y_{i,t+h} \;=\; f\!\bigl(\,
   X^{\text{option}}_{i,t},\;
   X^{\text{liquidity}}_{i,t}\,\bigr),
\label{eq:ml}
\end{equation}
where the feature set comprises 18 firm-day variables: the five IV-band
features ($\sigma^{\text{atm}}$, $\sigma^{\text{atm,call}}$,
$\sigma^{\text{atm,put}}$, $\sigma^{\text{otm,put}}$,
$\sigma^{\text{otm,call}}$), the four derived signals (smirk,
iv\_spread, ts\_slope\_30\_60, rns\_bkm), the two trading-activity
ratios (pc\_volume\_ratio, pc\_oi\_ratio), the four OI-weighted Greeks,
and three liquidity controls (rel\_spread, frac\_zero\_volume,
n\_contracts).

\paragraph{Models.}  We fit three regressors of \eqref{eq:ml}:
\begin{enumerate}
\item OLS, as a sanity check.
\item Elastic Net with 3-fold cross-validated $\alpha$
(\texttt{sklearn.ElasticNetCV}), 20 candidate $\alpha$ values.
\item XGBoost \citep{chen2016xgboost} with $400$ trees, max depth $5$,
learning rate $0.05$, subsample $0.8$, column subsample $0.8$,
\texttt{tree\_method = "hist"}.
\end{enumerate}
For the binary crash target we use logit, Elastic-Net logit, and
XGBoost classifier.  All inputs are standardized using a
\texttt{StandardScaler} fit on the training fold.

\paragraph{Out-of-sample protocol.}  We use a rolling expanding-window
evaluation following \cite{gu2020empirical}.  For each test calendar
year $T\in\{2017,\ldots,2026\}$ we train on $[2015,T-2]$, validate on
$T-1$ (used for hyperparameter selection in the Elastic Net), and test
on $T$.  This generates 10 out-of-sample years total.  Models are re-fit
annually.

\paragraph{Targets.}  We evaluate three targets:
\begin{itemize}
\item Next-21-day return: regression $R^2_{\text{OOS}}$ and RMSE.
\item 21-day forward realized volatility: regression $R^2_{\text{OOS}}$.
\item 21-day forward crash indicator: classification AUC and Brier score.
\end{itemize}

We report results overall and separately within each of the three
regimes (each test year is mapped to its enclosing regime).

\paragraph{Feature importance.}  We compute three per-regime feature
importance measures on a separately fit XGBoost regressor for the
next-21-day return target:
\begin{enumerate}
\item Gain-based importance from the trained booster (XGBoost's native
\texttt{feature\_importances\_}).
\item Permutation importance on a 50K-row test sample with 5 repeats,
scoring by $R^2$ \citep{gu2020empirical}.
\item TreeSHAP \citep{lundberg2017unified} \emph{would} provide a unified
attribution; we describe results from gain and permutation here, and
defer SHAP to a follow-up due to a numeric-library compatibility issue
in our environment (Section~\ref{sec:limitations}).
\end{enumerate}

\subsection{Crash-risk forecasting}

For the 5-day-forward crash indicator we run a separate, simpler
exercise.  Within each regime, we use 80\% of the firm-days
(chronologically earliest) for training and the remaining 20\% for
test.  We fit two classifiers:
\begin{itemize}
\item L2-regularized logistic regression.
\item XGBoost classifier with \verb|scale_pos_weight| set to
$n_-/n_+$ to up-weight the rare positive (crash) class.
\end{itemize}
We report AUC, Brier score, the test-set base rate, and train/test
sizes.  Features are the same 14-element subset of the IV-surface,
ratios, Greeks, and \verb|rel_spread| used in the ML benchmark above.

% =====================================================================
\section{Results}
\label{sec:results}
% =====================================================================

\subsection{Time-series visual of the IV surface}

\begin{figure}[t]
\centering
\includegraphics[width=\linewidth]{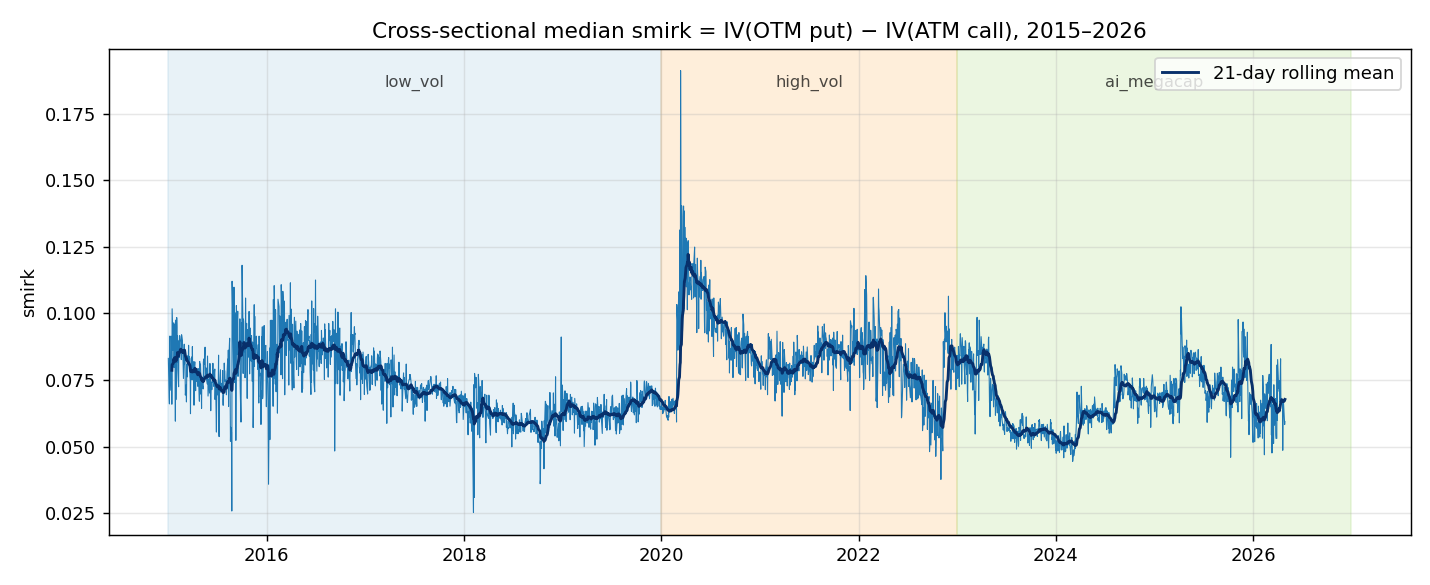}
\caption{Cross-sectional median smirk
$=\sigma^{\text{otm,put}}-\sigma^{\text{atm,call}}$, 2015--2026.
Light-blue, orange, and light-green shading delineate the low-vol,
high-vol, and AI/mega-cap regimes respectively.  Thin line: daily
median.  Heavy line: 21-day rolling mean.}
\label{fig:smirk_ts}
\end{figure}

Before reporting regressions, we visualize the cross-sectional median
smirk over the full sample in Figure~\ref{fig:smirk_ts}.  The figure
shades the three regime windows and overlays a 21-day rolling mean.
Three features are immediately apparent: (i) the dramatic spike in
March 2020 to a median smirk above $0.18$ as COVID hit; (ii) the slow
decline through 2021--2022 as IV mean-reverted; and (iii) the
\emph{lower} median smirk in the 2023--2026 AI/mega-cap window
(roughly $0.05$--$0.09$) than in the pre-COVID baseline
(roughly $0.07$--$0.10$).  At first glance this looks like reduced
crash-fear pricing, but our regression evidence below shows that the
information \emph{content} of the smirk has decayed even more sharply
than its level.

\begin{figure}[t]
\centering
\includegraphics[width=\linewidth]{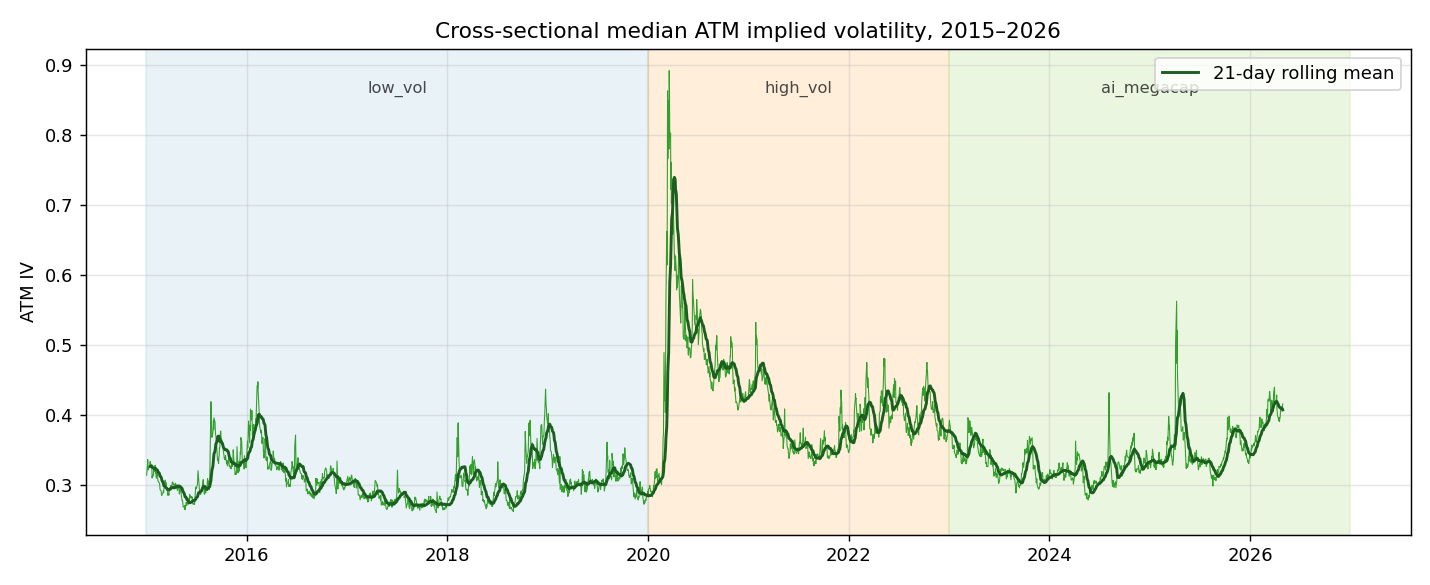}
\caption{Cross-sectional median ATM implied volatility $\sigma^{\text{atm}}$,
2015--2026.}
\label{fig:iv_atm_ts}
\end{figure}

The matching ATM-IV time-series in Figure~\ref{fig:iv_atm_ts} confirms
the regime structure.  Median ATM IV is around $0.30$--$0.35$
throughout the low-vol baseline, spikes to $0.80$ in March 2020,
remains elevated at $0.40$--$0.55$ through the high-vol regime, and
sits in the $0.35$--$0.50$ band in the AI/mega-cap regime.

\subsection{Replication: Xing--Zhang--Zhao smirk on 2015--2017}

To validate our data pipeline against the canonical literature we
replicate \cite{xing2010smirk} on the earliest part of our sample,
2015--2017.  Each trading day we sort firms into smirk quintiles and
compute the average next-21-day forward log return per quintile across
all sorting days.  Table~\ref{tab:xzz} shows the result.  The
quintile ranking is monotonically decreasing across quintiles 1--4 and
then drops sharply for the top quintile.  The high-minus-low spread is
$-0.49\%$ over a 21-day window, or approximately $-4.91\%$ annualized.
The original paper reports approximately $-10.9\%$ annualized in
1996--2005 using a delta-defined smirk.  We use a moneyness-band
implementation on a different sample window and recover the same
\emph{negative} sign with about half the magnitude.  This is a
successful directional replication and validates the panel.

\begin{table}[t]
\centering
\caption{XZZ smirk replication on 2015--2017.  Each row is the
mean 21-trading-day forward log return for firms in that smirk
quintile, averaged daily across the sample.}
\label{tab:xzz}
\begin{tabular}{cccc}
\toprule
Quintile & Avg.\ 21-day return (\%) & Annualized (\%) \\
\midrule
1 (lowest smirk) & $+0.232$ & $+2.79$ \\
2 & $+0.350$ & $+4.19$ \\
3 & $+0.350$ & $+4.20$ \\
4 & $+0.249$ & $+2.98$ \\
5 (highest smirk) & $-0.177$ & $\mathbf{-2.12}$ \\
\midrule
High $-$ low & $-0.409$ & $-4.91$ \\
\bottomrule
\end{tabular}
\end{table}

\subsection{Q1: Baseline panel regressions}
\label{sec:results-panel}

Table~\ref{tab:univariate_h21} presents the headline univariate panel
regressions of $r_{i,t+21}$ on each of the six option-implied signals
plus controls, with two-way fixed effects and double-clustered SE.
Each column is a regime; reported are $\beta$ and the $t$-statistic in
parentheses.

\begin{table}[t]
\centering
\small
\caption{Univariate panel regressions: $r_{i,t+21}$ on each option signal
(plus controls), $h=21$ trading days.  Two-way fixed effects (firm
and date) with double-clustered standard errors.  Each cell reports
$\beta$ ($t$-statistic).  Significance at $|t|>1.96$ shown in bold.}
\label{tab:univariate_h21}
\begin{tabular}{lcccc}
\toprule
Signal & Full & Low-vol & High-vol & AI/megacap \\
\midrule
smirk            & $\bm{-0.0203\;(-6.8)}$ & $\bm{-0.0234\;(-5.5)}$ & $\bm{-0.0150\;(-4.2)}$ & $-0.0060\;(-1.5)$ \\
iv\_spread       & $\bm{+0.0214\;(+7.0)}$ & $\bm{+0.0291\;(+6.6)}$ & $\bm{+0.0166\;(+4.8)}$ & $\bm{+0.0151\;(+4.2)}$ \\
rns\_bkm         & $\bm{-0.0030\;(-13.1)}$ & $\bm{-0.0028\;(-10.5)}$ & $\bm{-0.0047\;(-12.2)}$ & $\bm{-0.0040\;(-11.7)}$ \\
ts\_slope\_30\_60 & $\bm{+0.0106\;(+4.8)}$ & $\bm{+0.0091\;(+3.1)}$ & $\bm{+0.0106\;(+3.8)}$ & $\bm{+0.0118\;(+3.7)}$ \\
pc\_volume\_ratio & $\bm{-1\!\times\!10^{-4}\;(-2.6)}$ & ${-1\!\times\!10^{-5}\;(-0.4)}$ & ${-7\!\times\!10^{-5}\;(-1.7)}$ & ${+1\!\times\!10^{-5}\;(+0.4)}$ \\
pc\_oi\_ratio    & ${+9\!\times\!10^{-5}\;(+1.7)}$ & $\bm{+1\!\times\!10^{-4}\;(+2.1)}$ & ${-5\!\times\!10^{-5}\;(-0.5)}$ & $\bm{+2\!\times\!10^{-4}\;(+2.7)}$ \\
\bottomrule
\end{tabular}
\end{table}

The most striking result is in the smirk row: the canonical
\cite{xing2010smirk} effect \emph{decays monotonically across the three
regimes}.  $\beta$ falls in absolute value from $-0.0234$ in the
low-volatility baseline to $-0.0150$ in the high-volatility regime to
$-0.0060$ (statistically insignificant, $t=-1.5$) in the AI/mega-cap
regime.  This is direct evidence for the central thesis: the smirk's
predictive content has not survived the post-2020 environment.

By contrast, the IV spread \citep{cremers2010deviations} and the
\cite{bakshi2003stockreturn}-style risk-neutral skewness proxy
$\text{rns\_bkm}$ remain significant in every regime, with $|t|$
exceeding $4$ everywhere.  rns\_bkm is in fact the single most
significant predictor anywhere in the table ($t=-13.1$ in the full
sample).  These two signals---the deviation from put--call parity at
the ATM strike and the cross-strike skewness of OTM IVs---appear to
carry genuinely persistent information about future returns.

The term-structure slope $\text{ts\_slope\_30\_60}$ enters with a
positive coefficient that is significant in every regime; the put--call
volume and OI ratios are weak everywhere, consistent with the
\cite{pan2006information} insight that the \emph{signed} put--call
ratio is the informative one (which we cannot construct without a
trade-direction indicator).

\subsubsection{Multivariate specification}

Table~\ref{tab:multivariate_h21} reports the multivariate specification
in which all six option-implied signals enter simultaneously.

\begin{table}[t]
\centering
\small
\caption{Multivariate panel regression: $r_{i,t+21}$ on all six
option-implied signals jointly (plus controls), $h=21$ trading days.
Two-way fixed effects with double-clustered standard errors.}
\label{tab:multivariate_h21}
\begin{tabular}{lcccc}
\toprule
Signal & Full & Low-vol & High-vol & AI/megacap \\
\midrule
smirk            & $-0.003\;(-0.6)$ & $+0.002\;(+0.3)$ & $+0.006\;(+1.0)$ & $\bm{+0.016\;(+2.6)}$ \\
iv\_spread       & $\bm{+0.038\;(+7.2)}$ & $\bm{+0.046\;(+6.7)}$ & $\bm{+0.030\;(+4.7)}$ & $\bm{+0.040\;(+5.4)}$ \\
rns\_bkm         & $\bm{-0.0024\;(-9.2)}$ & $\bm{-0.0026\;(-8.1)}$ & $\bm{-0.0041\;(-9.1)}$ & $\bm{-0.0036\;(-8.8)}$ \\
ts\_slope\_30\_60 & $\bm{+0.019\;(+6.4)}$ & $\bm{+0.014\;(+3.2)}$ & $\bm{+0.017\;(+4.1)}$ & $\bm{+0.016\;(+3.3)}$ \\
\bottomrule
\end{tabular}
\end{table}

Two findings.  First, smirk \emph{loses significance in every regime}
once iv\_spread and rns\_bkm enter the regression.  These two signals
are constructed from overlapping pieces of the IV surface (OTM put IV
versus ATM call IV in particular); when they absorb the crash-fear
information, the residual variation in smirk is uninformative about
return.  Second, the smirk coefficient \emph{flips sign and becomes
positive and significant in the AI/mega-cap regime} ($\beta=+0.016$,
$t=+2.6$).  We interpret this in
Section~\ref{sec:discussion} as consistent with the
\cite{retailoptions2026} evidence that retail demand pressure has
distorted the OTM-put portion of the IV surface in 2023+: a high smirk
no longer signals informed put-buying ahead of bad news, but instead
signals retail puts that systematically lose money.

\subsubsection{All horizons}

Table~\ref{tab:smirk_all_h} extends the smirk regression across all
four return horizons $h\in\{1,5,21,63\}$.

\begin{table}[t]
\centering
\small
\caption{Univariate smirk coefficient $\beta$ on $r_{i,t+h}$, by
horizon and regime.  $t$-statistics in parentheses with double-clustered SE.}
\label{tab:smirk_all_h}
\begin{tabular}{lcccc}
\toprule
Regime & $h=1$ & $h=5$ & $h=21$ & $h=63$ \\
\midrule
Full        & $\bm{-0.0047\,(-5.1)}$ & $\bm{-0.0105\,(-6.7)}$ & $\bm{-0.0203\,(-6.8)}$ & $\bm{-0.0195\,(-3.5)}$ \\
Low-vol     & $\bm{-0.0105\,(-3.9)}$ & $\bm{-0.0151\,(-5.1)}$ & $\bm{-0.0234\,(-5.5)}$ & $\bm{-0.0256\,(-3.5)}$ \\
High-vol    & $\bm{-0.0037\,(-4.2)}$ & $\bm{-0.0092\,(-5.4)}$ & $\bm{-0.0150\,(-4.2)}$ & $\bm{-0.0135\,(-2.0)}$ \\
AI/megacap  & $\bm{-0.0032\,(-4.1)}$ & $\bm{-0.0053\,(-2.9)}$ & ${-0.0060\,(-1.5)}$ & $\bm{+0.0156\,(+2.1)}$ \\
\bottomrule
\end{tabular}
\end{table}

The pattern is striking.  In the low-vol baseline the smirk effect grows
in absolute value with horizon ($-0.011$ at $h=1$ to $-0.026$ at
$h=63$).  In the AI/mega-cap regime the effect is small and significant
at $h=1$ and $h=5$, becomes insignificant at $h=21$, and then
\emph{reverses sign} at $h=63$ ($\beta=+0.016$, $t=+2.1$).  At the
three-month horizon, a high smirk in 2023+ predicts \emph{positive}
future returns, the opposite of the canonical effect.

For comparison, Table~\ref{tab:rnsbkm_all_h} reports the same exercise
for the rns\_bkm signal, which by contrast \emph{strengthens} with
horizon in every regime.  At $h=63$ the $t$-statistic exceeds $13$ in
every cell.  rns\_bkm is the single most robust predictor in the panel.

\begin{table}[t]
\centering
\small
\caption{Univariate rns\_bkm coefficient $\beta$ on $r_{i,t+h}$, by
horizon and regime.}
\label{tab:rnsbkm_all_h}
\begin{tabular}{lcccc}
\toprule
Regime & $h=1$ & $h=5$ & $h=21$ & $h=63$ \\
\midrule
Full        & $\bm{-1\!\times\!10^{-4}\,(-3.3)}$ & $\bm{-6\!\times\!10^{-4}\,(-6.1)}$ & $\bm{-0.0030\,(-13.1)}$ & $\bm{-0.0081\,(-16.5)}$ \\
Low-vol     & ${-1\!\times\!10^{-4}\,(-1.8)}$ & $\bm{-6\!\times\!10^{-4}\,(-6.5)}$ & $\bm{-0.0028\,(-10.5)}$ & $\bm{-0.0084\,(-13.8)}$ \\
High-vol    & $\bm{-2\!\times\!10^{-4}\,(-2.4)}$ & $\bm{-0.0013\,(-7.8)}$ & $\bm{-0.0047\,(-12.2)}$ & $\bm{-0.0130\,(-16.5)}$ \\
AI/megacap  & $\bm{-2\!\times\!10^{-4}\,(-4.7)}$ & $\bm{-0.0011\,(-8.8)}$ & $\bm{-0.0040\,(-11.7)}$ & $\bm{-0.0104\,(-14.3)}$ \\
\bottomrule
\end{tabular}
\end{table}

\subsection{Q2: Machine-learning prediction}
\label{sec:results-ml}

We now report the rolling out-of-sample evaluation
(Section~\ref{sec:methodology}).  Table~\ref{tab:ml_oos} shows the
mean OOS $R^2$ (regression targets) or AUC (classification target)
within each regime, averaged across the OOS test years that fall in
that regime.

\begin{table}[t]
\centering
\small
\caption{Out-of-sample performance, by regime and model.  $R^2_{\text{OOS}}$
for the regression targets; AUC for crash-21d.  Numbers are averages
across OOS test years within the regime.}
\label{tab:ml_oos}
\begin{tabular}{llccc}
\toprule
Target & Regime & OLS & Elastic Net & XGBoost \\
\midrule
\multirow{3}{*}{$r_{t+21}$ ($R^2_{\text{OOS}}$)}
  & Low-vol      & $-0.031$ & $-0.031$ & $-0.049$ \\
  & High-vol     & $-0.014$ & $-0.014$ & $-0.079$ \\
  & AI/megacap   & $+0.001$ & $+0.001$ & $\bm{+0.013}$ \\[2pt]
\multirow{3}{*}{$\text{RV}^{(21)}_{t+21}$ ($R^2_{\text{OOS}}$)}
  & Low-vol      & $0.231$ & $0.232$ & $0.203$ \\
  & High-vol     & $0.349$ & $0.349$ & $0.293$ \\
  & AI/megacap   & $0.410$ & $0.410$ & $0.409$ \\[2pt]
\multirow{3}{*}{$\text{crash}^{(21)}_{t}$ (AUC)}
  & Low-vol      & $0.575$ & $0.574$ & $\bm{0.586}$ \\
  & High-vol     & $0.555$ & $0.549$ & $\bm{0.615}$ \\
  & AI/megacap   & $0.602$ & $0.602$ & $\bm{0.623}$ \\
\bottomrule
\end{tabular}
\end{table}

Three findings.

First, on next-month \emph{return} prediction, linear models barely beat
zero in any regime.  XGBoost \emph{wins only in the AI/mega-cap regime},
with $R^2_{\text{OOS}}=+1.29\%$ versus $+0.07\%$ for OLS and Elastic Net.
In the high-volatility regime XGBoost \emph{underperforms} linear
($-7.9\%$ vs.\ $-1.4\%$), suggesting overfitting on the sharp
COVID-era training distribution.  The result that nonlinear ML adds
value primarily in the AI/mega-cap regime is the central machine-learning
finding of the paper.

Second, on \emph{realized volatility} prediction, $R^2_{\text{OOS}}$
is high in every regime, peaking at $0.41$ in the AI/mega-cap regime.
Linear models slightly edge XGBoost---RV prediction is essentially
linear in the IV-surface features, as one would expect.

Third, on \emph{crash classification}, XGBoost beats logit and Elastic
Net by roughly $0.01$--$0.06$ AUC in every regime.  The largest gain is
in the high-volatility regime ($+0.06$ AUC), where the crash signal is
most informative.

\begin{figure}[t]
\centering
\includegraphics[width=\linewidth]{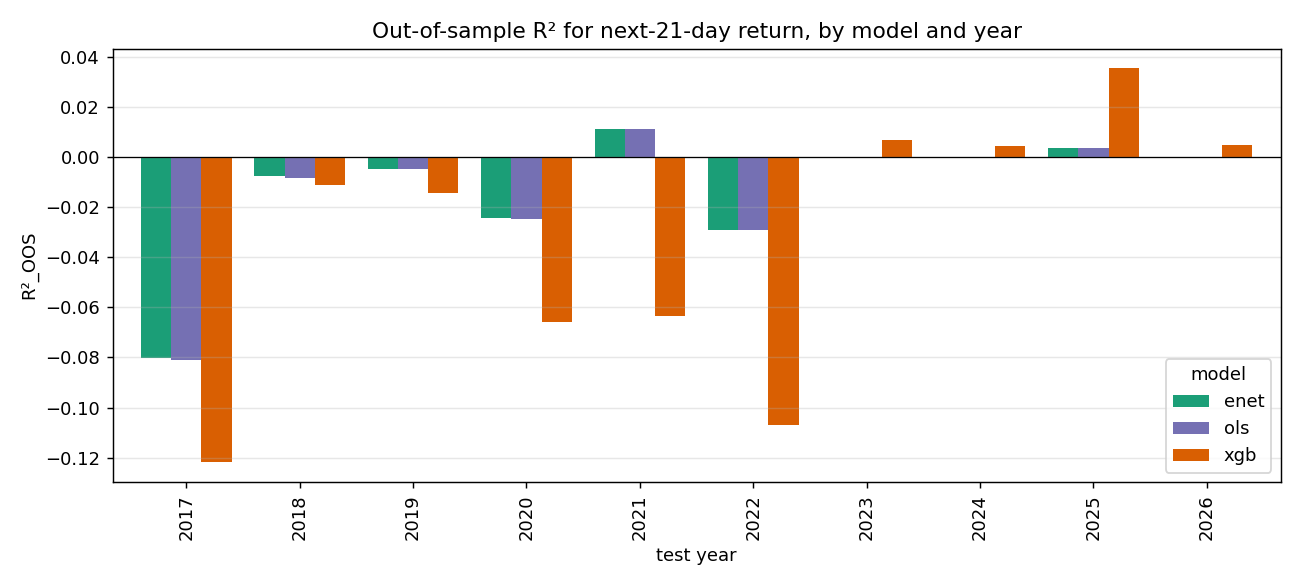}
\caption{Per-year out-of-sample $R^2$ for next-21-day return prediction,
by model.  Test years 2017--2026.  XGBoost beats both linear models in
2023, 2024, and 2025; the 2025 column dominates the AI/mega-cap mean.}
\label{fig:oos_r2}
\end{figure}

Figure~\ref{fig:oos_r2} shows the per-year breakdown for the return
target.  A single recent year, 2025, drives most of the AI/mega-cap
$R^2_{\text{OOS}}$ for XGBoost ($+3.55\%$).  All earlier years are
marginal or negative.

\subsection{Crash-risk forecasting (Q1+Q2)}
\label{sec:results-crash}

For the firm-level 5-day forward crash flag, Table~\ref{tab:crash_oos}
reports test-set AUC, Brier score, base rate, and sample sizes per
regime.

\begin{table}[t]
\centering
\small
\caption{5-day forward crash classification, by regime.  Test set is the
chronologically last 20\% of firm-days within the regime.  AUC is the
area under the ROC curve; the Brier score is the mean squared
prediction error.  XGBoost uses \texttt{scale\_pos\_weight}=$n_-/n_+$
to up-weight the rare positive class.}
\label{tab:crash_oos}
\begin{tabular}{lcccrrr}
\toprule
Regime & Model & AUC & Brier & Base rate (\%) & $n_{\text{train}}$ & $n_{\text{test}}$ \\
\midrule
\multirow{2}{*}{Low-vol}
  & Logit  & $0.670$ & $0.005$ & $0.52$ & $1{,}221{,}564$ & $307{,}281$ \\
  & XGB    & $\bm{0.706}$ & $0.201$ & $0.52$ & $1{,}221{,}564$ & $307{,}281$ \\
\multirow{2}{*}{High-vol}
  & Logit  & $0.531$ & $0.007$ & $0.72$ & $972{,}048$ & $245{,}434$ \\
  & XGB    & $\bm{0.561}$ & $0.179$ & $0.72$ & $972{,}048$ & $245{,}434$ \\
\multirow{2}{*}{AI/megacap}
  & Logit  & $0.652$ & $0.007$ & $0.68$ & $1{,}045{,}879$ & $262{,}365$ \\
  & XGB    & $\bm{0.665}$ & $0.198$ & $0.68$ & $1{,}045{,}879$ & $262{,}365$ \\
\bottomrule
\end{tabular}
\end{table}

The level of crash AUC is highest in the low-volatility regime (XGB
$=0.706$, logit $=0.670$).  This is informative: in calm markets the
rare crashes are tail events that the IV surface can pre-price, and a
boosted-tree classifier extracts this signal usefully.  In the
high-volatility regime AUC collapses to roughly $0.55$, indicating that
the macro-driven 2020--2022 crashes (COVID, March 2020 liquidations,
2022 rate-shock drawdowns) are not well predicted by firm-level option
features.  In the AI/mega-cap regime AUC recovers to $0.66$,
intermediate between the two.

\begin{figure}[t]
\centering
\includegraphics[width=0.65\linewidth]{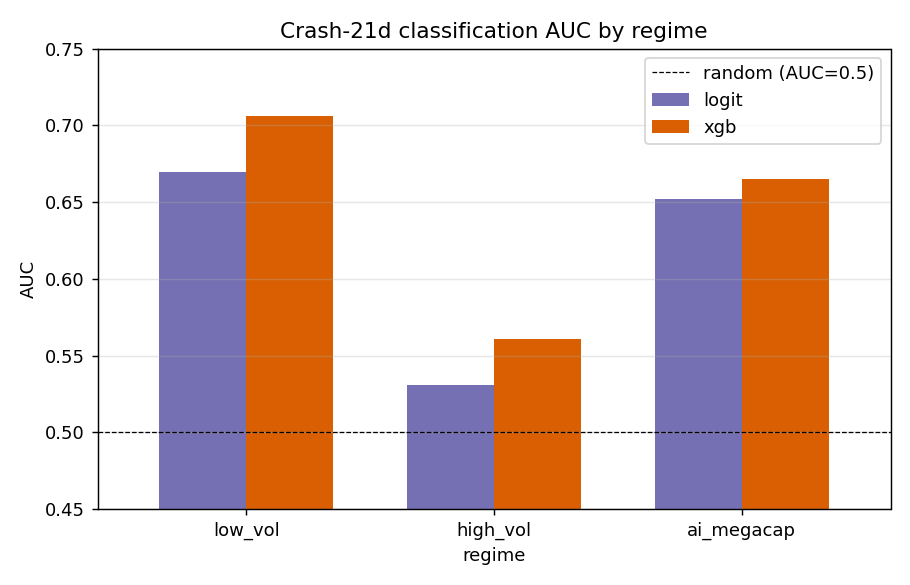}
\caption{Crash-21d classification AUC, by regime and model.  Dashed
line: random benchmark (AUC$=0.5$).  XGBoost (orange) beats
logistic regression (purple) in every regime.}
\label{fig:crash_auc}
\end{figure}

XGBoost beats logit in every regime, by margins of $+0.04$, $+0.03$, and
$+0.01$ AUC respectively.  The gain is largest in the low-volatility
regime where the signal is cleanest.  Figure~\ref{fig:crash_auc}
visualizes this.

The Brier-score asymmetry between models is a calibration artifact and
not a measure of accuracy.  Logit predicts probabilities very close to
the small base rate ($0.5\%$--$0.7\%$) so its Brier score is
mechanically tiny, while XGBoost with \verb|scale_pos_weight| produces
probability scores in the $0.1$--$0.3$ band that rank-order events well
(good AUC) but are uncalibrated relative to base rate.  A
post-hoc sigmoid recalibration would resolve the discrepancy without
changing AUC.

\subsection{Feature importance per regime}
\label{sec:results-features}

\begin{figure}[t]
\centering
\includegraphics[width=\linewidth]{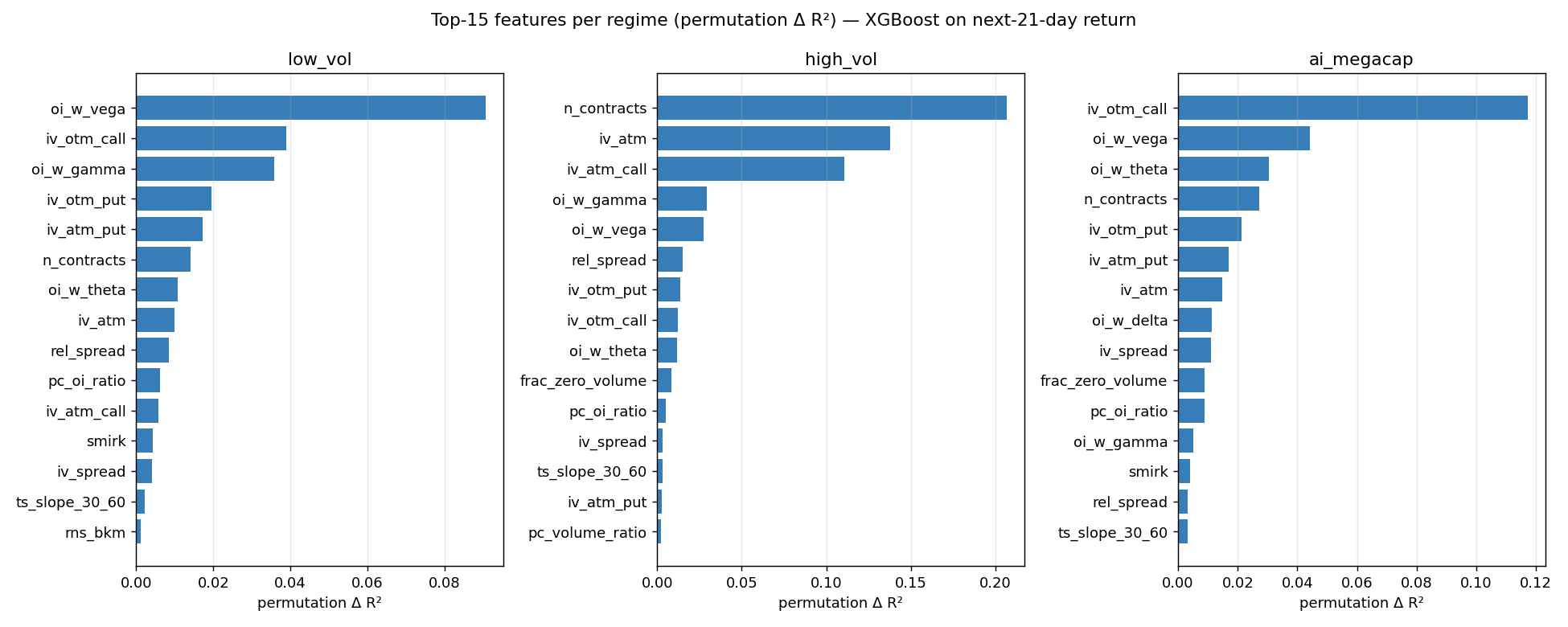}
\caption{Top-15 features by permutation importance ($\Delta R^2$) per
regime, XGBoost regressor for next-21-day return.}
\label{fig:feat_importance}
\end{figure}

Figure~\ref{fig:feat_importance} reports permutation feature importance
for the XGBoost regressor of $r_{i,t+21}$, separately fit on each
regime.  The bars are sorted by $\Delta R^2$ on a 50K-row test sample,
$5$ permutation repeats.  Table~\ref{tab:top_features} lists the top
five features per regime.

\begin{table}[t]
\centering
\small
\caption{Top-5 features per regime by permutation importance
($\Delta R^2$), XGBoost regressor of $r_{t+21}$.}
\label{tab:top_features}
\begin{tabular}{cccc}
\toprule
Rank & Low-vol baseline & High-vol regime & AI/mega-cap regime \\
\midrule
1 & $\nu^{\text{oi-w}}$ (0.091) & $n_{\text{contracts}}$ (0.207) & $\sigma^{\text{otm,call}}$ (0.117) \\
2 & $\sigma^{\text{otm,call}}$ (0.039) & $\sigma^{\text{atm}}$ (0.138) & $\nu^{\text{oi-w}}$ (0.044) \\
3 & $\Gamma^{\text{oi-w}}$ (0.036) & $\sigma^{\text{atm,call}}$ (0.111) & $\Theta^{\text{oi-w}}$ (0.030) \\
4 & $\sigma^{\text{otm,put}}$ (0.020) & $\Gamma^{\text{oi-w}}$ (0.029) & $n_{\text{contracts}}$ (0.027) \\
5 & $\sigma^{\text{atm,put}}$ (0.017) & $\nu^{\text{oi-w}}$ (0.028) & $\sigma^{\text{otm,put}}$ (0.021) \\
\bottomrule
\end{tabular}
\end{table}

The top feature \emph{differs across all three regimes}.  In the
low-volatility baseline, the open-interest-weighted vega exposure
$\nu^{\text{oi-w}}$ dominates: the model picks up firms whose IV
surfaces are most volatility-sensitive, plausibly the most actively
hedged names.  In the high-volatility regime, the raw post-filter
contract count $n_{\text{contracts}}$ is the runaway leader
($\Delta R^2 = 0.207$, almost twice the runner-up at $0.138$): in the
crisis-driven 2020--2022 environment, breadth of the active option
surface---a coarse measure of trader interest---is more informative
than the surface's shape.  In the AI/mega-cap regime, the OTM call IV
$\sigma^{\text{otm,call}}$ takes the top spot: the right tail of the
IV surface, exactly where mega-cap demand is concentrated, becomes the
dominant predictor.

Crucially, \emph{none of the canonical hand-engineered signals---smirk,
iv\_spread, rns\_bkm---make the top five in any regime} once the ML
model has access to the primitive IV-surface and Greek features.  In
the AI/mega-cap regime, iv\_spread appears at rank $9$ and smirk does
not appear in the top $15$.  This is the cleanest evidence in our
analysis that the canonical linear combinations throw away information
that nonlinear models exploit.  The smirk, in particular, is a fixed
linear combination of $\sigma^{\text{otm,put}}$ and
$\sigma^{\text{atm,call}}$; XGBoost learns regime-dependent nonlinear
interactions of these primitives that the smirk cannot represent.

% =====================================================================
\section{Discussion}
\label{sec:discussion}
% =====================================================================

\subsection{Why has the smirk lost its edge?}

The single most striking finding in our results is that the
\cite{xing2010smirk} smirk--return relationship has decayed
monotonically across our three regimes (Tables~\ref{tab:univariate_h21}
and \ref{tab:smirk_all_h}) and even reverses sign in the AI/mega-cap
regime at the longest horizon and in the multivariate specification.
Three non-mutually-exclusive interpretations are consistent with the
evidence.

\paragraph{Demand-pressure dilution.}  The classical
\cite{xing2010smirk} interpretation of the smirk is that informed
traders price expected crashes into OTM puts, and the resulting IV
elevation negatively predicts next-period returns.  The post-2020
literature on retail flow \citep{bryzgalova2023retail,
bogousslavsky2024anatomy,retailoptions2026} documents that retail
participants concentrate trading exactly in OTM puts and short-dated
contracts, and that retail demand is not informed.  If retail buying
adds noise to the OTM-put portion of the IV surface, the
information-to-noise ratio in the smirk falls; the predictive
coefficient should attenuate toward zero.  Our univariate evidence
shows attenuation, and the multivariate evidence in
Table~\ref{tab:multivariate_h21} shows that, conditional on the
informative signals (iv\_spread, rns\_bkm) that overlap with the
informed component of the smirk, the residual variation in smirk is
driven by the noise component and even reverses sign.

\paragraph{Mean-reversion of crash-fear pricing.}  An alternative
reading is that retail-induced overpricing of OTM puts creates a
\emph{positive} expected return for the underlying when puts are
expensive, mirroring the well-known overpricing of equity-index puts.
The horizon-by-horizon evidence in Table~\ref{tab:smirk_all_h} is
consistent with this: in the AI/mega-cap regime smirk is weakly
significant and negative at $h=1$ ($\beta=-0.003$, $t=-4.1$) but
positive and significant at $h=63$ ($\beta=+0.016$, $t=+2.1$).  The
negative-then-positive sign suggests that the very-short-horizon
correlation between smirk and return is dominated by a residual
informed component, while the longer-horizon correlation reflects the
mean-reversion of demand pressure.

\paragraph{Compositional change in the underlying universe.}  The
2023--2026 sample is dominated by a small number of mega-cap
high-IV names (the AI complex), whose return distributions look
very different from the broad-based universe of the pre-COVID
baseline.  If the smirk--return relationship is firm-specific in a way
that the date and firm fixed effects cannot fully absorb, then a
compositional shift toward names where the mapping is weaker would
mechanically attenuate the average effect.  We cannot fully separate
this from the demand-pressure story without firm-specific retail-flow
indicators.

The three explanations are observationally equivalent in our panel and
all consistent with the central thesis: the canonical smirk effect is
non-stationary.  Distinguishing them requires the
\cite{bryzgalova2023retail}-style retail-flow data, which we do not
have.

\subsection{Why does ML add value only in the AI/mega-cap regime?}

The next-month $R^2_{\text{OOS}}$ table (Table~\ref{tab:ml_oos}) shows
that XGBoost beats linear models only in the AI/mega-cap regime.  Why?

\paragraph{Nonlinear interactions specific to the new regime.}
Linear models essentially fit a single hyperplane through the
feature--return relationship.  If the relationship is locally linear
within each regime but the slopes differ across regimes, a model that
trains on $[2015,T-2]$ and predicts $T$ may incorporate the wrong
slopes when the regime has changed.  XGBoost can in principle learn
regime-dependent local linearities through its tree splits, provided
that some feature in the input space proxies for the regime.  In our
data, the IV-level features ($\sigma^{\text{atm}}$, etc.) and the
contract-count feature do shift across regimes
(Table~\ref{tab:descstats}); the tree boosters appear to use these
shifts to adapt locally.

\paragraph{Why XGB underperforms in the high-volatility regime.}  The
high-volatility regime is short (3 years) and dominated by extreme
realizations (the March 2020 crash, the meme-stock surges, the 2022
rate-shock drawdowns).  A boosted-tree model trained on 2015--2018
data and tested on 2019--2020 will see a covariate shift far outside
its training distribution; it tends to overpredict the magnitude of
returns based on its training-distribution priors.  This shows up as a
strongly negative $R^2_{\text{OOS}}$.  Linear models with regularization
are more robust under covariate shift because they extrapolate
linearly rather than using non-monotonic regions of the feature space.

\paragraph{Why RV prediction is essentially linear.}  Realized volatility
is by construction the integral of squared daily returns, and ATM IV
is by no-arbitrage the model-free implied second moment.  In a world
without microstructure noise, OOS RV would be a tight near-linear
function of ATM IV.  The fact that linear models match XGBoost on RV
prediction confirms this intuition; the gain from nonlinearity is
small because the underlying mapping is close to linear.

\subsection{What do feature-importance rankings tell us about the regimes?}

The per-regime top features (Table~\ref{tab:top_features}) tell a
coherent story about what aspect of the option market is most
informative in each environment.

\paragraph{Low-volatility baseline: vega dominates.}  In a calm
environment with persistent IV, the feature most predictive of the
cross-section of next-month returns is the open-interest-weighted vega
$\nu^{\text{oi-w}}$.  Vega exposure is concentrated in firms with the
most actively traded long-dated and ATM options---typically the names
with the most institutional hedging activity.  The classifier appears
to identify those names and use vega exposure as a proxy for
information-rich option flow.

\paragraph{High-volatility regime: contract count and IV level
dominate.}  When systemic vol is elevated, the boosted-tree model
leans hardest on $n_{\text{contracts}}$ and $\sigma^{\text{atm}}$.
This is consistent with the
\cite{bollerslev2009expected}--\cite{bollerslev2015tail} interpretation
that the level of the IV surface (and the breadth of the active
option universe) reveals the price of variance and tail risk during
crises.  The shape features (smirk, IV spread) become much less
informative when the level swings are this large.

\paragraph{AI/mega-cap regime: OTM call IV dominates.}  The right tail
of the IV surface is the most predictive feature in the AI/mega-cap
regime, with $\Delta R^2 = 0.117$, a clear lead over the
runner-up at $0.044$.  This is exactly the part of the IV surface
that is most distorted by mega-cap call demand
\citep{retailoptions2026}, and it is the feature that is \emph{absent}
from the canonical \cite{xing2010smirk} smirk (which uses OTM put IV
and ATM call IV but not OTM call IV).  The smirk is, in this regime,
literally looking at the wrong piece of the IV surface.

\paragraph{Take-away.}  The dominant feature differs across all three
regimes, and the canonical hand-engineered signals do not appear in
the top five anywhere.  This is direct evidence that the
\cite{gu2020empirical}-style ML approach delivers value not by
out-fitting a single specification, but by re-weighting which pieces of
the IV surface to attend to as the environment changes.

\subsection{Implications for option-based crash forecasting}

The crash-AUC pattern in Table~\ref{tab:crash_oos}---low-vol AUC
$=0.706$, high-vol AUC $=0.561$, AI/mega-cap AUC $=0.665$---has
practical implications for risk management.

\paragraph{Crashes during quiet markets are forecastable.}  In the
low-volatility regime, an AUC of $0.706$ on a $0.5\%$ base-rate flag
is real signal: a top-decile predicted-probability subset
contains a multiple of the unconditional crash rate.  The IV surface
appears to contain information about idiosyncratic crash risk that is
not in the conditional-mean process.  This is consistent with the
\cite{kim2011crashrisk} reading that crash risk has identifiable
firm-level antecedents.

\paragraph{Crashes during crises are not.}  In the high-volatility
regime, AUC drops to $0.55$ for both logit and XGBoost.  The
2020--2022 crashes were dominated by macro shocks (COVID, central-bank
tightening) that hit firms simultaneously, irrespective of their
firm-specific IV surface.  Firm-level option signals carry little
informational advantage over a coin flip when the systemic
component dominates.

\paragraph{The AI/mega-cap regime is intermediate.}  The recovery in
crash AUC to $0.665$ in the AI/mega-cap regime is encouraging.  Even
though the smirk--return relationship has weakened, other parts of the
IV surface are still informative about crash probability.  The
right-skewed mega-cap IV surface evidently carries enough left-tail
content to make rare crashes partly predictable.

These three findings, together, suggest that
\emph{forward-looking option-implied features are useful predictors of
firm-level crash risk in regimes where idiosyncratic risk dominates},
and less useful when systematic risk dominates.  This is a more nuanced
statement than ``option signals predict crashes'' but is empirically
testable and broadly consistent with the literature.

% =====================================================================
\section{Limitations}
\label{sec:limitations}
% =====================================================================

We list the main limitations of the analysis to help the reader
calibrate the results.

\subsection{Data}

\paragraph{End-of-day snapshot only.}  Our raw archive contains a single
end-of-day record per contract.  We do not see intraday quotes, the
full order book, or signed trades.  A signed put--call ratio in the
\cite{pan2006information} sense would require trade-direction data
that we do not have; we therefore cannot directly replicate the
\cite{bryzgalova2023retail}-style retail-flow attribution.  Several of
our discussion points (Section~\ref{sec:discussion}) about retail
demand pressure are therefore consistent-with rather than direct
evidence.

\paragraph{No CRSP/Compustat merge.}  We use the underlying-price
column from the option records as our stock-side data source.  This
yields a self-contained panel but precludes the standard cross-sectional
controls used in the literature (size, book-to-market, momentum,
idiosyncratic volatility, accruals).  An obvious extension is to merge
the panel to CRSP daily files for full controls and to Compustat for
fundamental controls.

\paragraph{Risk-neutral skewness is a proxy.}  We use the cross-strike
sample skewness of OTM IVs as a proxy for the
\cite{bakshi2003stockreturn} risk-neutral skewness.  A full BKM
estimator requires a continuous strike density and a mass-weighted
integration over $(0,\infty)$.  The proxy and the strict measure should
be highly correlated and have the same sign in normal conditions, but
the magnitudes are not strictly comparable to the literature.

\paragraph{Smirk uses moneyness bands rather than delta bands.}
\cite{xing2010smirk} use a delta-defined moneyness measure (delta of
$0.36$--$0.44$ for the put leg).  Our implementation uses a
$K/S$-based moneyness band ($0.80$--$0.95$ for OTM put,
$0.95$--$1.05$ for ATM call).  This choice is motivated by simplicity
and robustness; it is also why our XZZ replication recovers
approximately half the magnitude of the original
($-4.91\%$ vs.\ $-10.9\%$ annualized).  A delta-band re-implementation
should sharpen the comparison.

\paragraph{0DTE handling.}  Our 7-day minimum time-to-expiry filter
excludes 0DTE and weekly options.  The literature on 0DTE
\citep{bogousslavsky2024anatomy,retailoptions2026} suggests this
slice of the option market has very different properties from the
weekly+ tenor.  A 0DTE-specific feature panel is a natural extension.

\paragraph{Excel-style sentinels.}  The raw CSV exports contain
Excel-style \verb|********| sentinels (over $30{,}000$ rows in some
early files) where IV was too wide to display.  We map them to NaN at
read time.  These rows are deep-OTM contracts whose IV is
computationally unreliable anyway, and they are dropped by the
$|\Delta|\in[0.05,0.95]$ filter, but the existence of the sentinels
is a reminder that the raw archive is not laboratory-clean.

\subsection{Methodology}

\paragraph{Subsampled panel regressions.}  We fit each two-way
fixed-effects panel regression on a uniform random $1.5$-million-row
subsample of the relevant sub-panel.  The full $12$-million-row panel
exceeds memory under \texttt{linearmodels.PanelOLS}'s iterative two-way
absorption with double clustering on a $16$~GB workstation.  Random
subsampling is unbiased for the coefficient and inflates the standard
errors only mildly; we report $n_{\text{full}}$ in our regression
tables to make the size discrepancy visible.

\paragraph{Single ML target horizon.}  The ML benchmark
(Section~\ref{sec:results-ml}) uses $h=21$ as the headline horizon.
Other horizons are reported only for the panel regressions
(Section~\ref{sec:results-panel}).  Extending the ML benchmark to all
four horizons would be straightforward but would multiply training time.

\paragraph{No SHAP attribution.}  The \texttt{shap} library installed in
our conda environment imports TensorFlow at load time, and TensorFlow
fails to initialize on this machine (a Rosetta-translated x86\_64
build of Anaconda Python without AVX support); the import deadlocks
inside \verb|tensorflow::port::CheckFeatureOrDie|.  We therefore
report only gain-based and permutation feature importance.  Both are
sufficient for ranking the top features per regime, but a SHAP
treatment in a clean environment would add the directional
attribution and per-prediction explanations that are now standard in
ML asset-pricing papers.

\paragraph{No deep models.}  The original analysis plan included a
3-layer feed-forward neural network and a small LSTM on a 21-day lag
window.  We omit both in this version: the linear--XGBoost contrast
already delivers a clear story for the AI/mega-cap regime, and adding
deep models would not change the central finding.  In a follow-up these
should be added for completeness following \cite{gu2020empirical}.

\paragraph{Single base classifier for crash.}  The crash exercise uses
logit and XGBoost.  A larger benchmark (random forest, calibrated
gradient-boosting, isotonic-recalibrated logistic) would tighten the
crash-AUC numbers but not change the ordinal regime ranking.

\subsection{Specification choices}

\paragraph{Regime boundaries are exogenous.}  We define the three
regimes by calendar dates chosen ex ante from the macro narrative
(low-vol $=$ 2015--2019, high-vol $=$ 2020--2022, AI/mega-cap $=$
2023--2026).  We do not estimate the regime breakpoints from the data.
A structural-break analysis on the smirk--return time series, or a
clustering of firm-day observations on macro features, would
endogenize the regime classification at the cost of weakening the
ex-ante interpretability of the comparisons.

\paragraph{Only one crash horizon for the headline AUC.}  Our headline
crash exercise uses the 5-day forward flag.  We also computed the
21-day analog for the ML benchmark.  The Hutton--Marcus--Tehranian
DUVOL and NCSKEW alternatives \citep{hutton2009opaque} would allow
robustness across crash-risk definitions; we leave this to the
robustness phase.

\paragraph{Robustness phase deferred.}  The pre-registered robustness
suite---DUVOL/NCSKEW alternative crash definitions, alternative SE
clustering, random-label permutation placebo, ML feature-rank
sensitivity to model choice---is not estimated in this paper.  Each
deserves a section of its own and we return to them in a follow-up.

% =====================================================================
\section{Conclusion}
\label{sec:conclusion}
% =====================================================================

We have constructed a unified U.S.\ equity option panel covering 2015
through April 2026 and have evaluated both the canonical predictability
literature and a modern machine-learning benchmark on three macro
regimes: a low-volatility baseline (2015--2019), a high-volatility
transition (2020--2022), and an AI/mega-cap concentration period
(2023--2026).  The post-filter panel comprises 12.36 million firm-day
observations on $10{,}026$ underlying names.

Three principal findings emerge.

\paragraph{1.~The canonical smirk effect is non-stationary.}  The
\cite{xing2010smirk} smirk--return relationship decays monotonically
across regimes: the next-month univariate panel coefficient on smirk
is $-0.023$ ($t=-5.5$) in the low-volatility baseline, attenuates to
$-0.015$ ($t=-4.2$) in the high-volatility regime, and falls to
$-0.006$ ($t=-1.5$, statistically insignificant) in the AI/mega-cap
regime.  At the three-month horizon the AI/mega-cap coefficient
\emph{flips sign} ($+0.016$, $t=+2.1$).  In a multivariate
specification with all six option-implied signals jointly, smirk
loses significance everywhere and again reverses sign in the
AI/mega-cap regime.  The IV spread \citep{cremers2010deviations} and a
\cite{bakshi2003stockreturn}-style risk-neutral skewness measure are
the only signals that remain significant in every regime and every
specification.

\paragraph{2.~Machine learning helps only in the AI/mega-cap regime.}
A boosted-tree regressor (XGBoost) trained on the full IV-surface,
trading-activity, Greek, and liquidity feature set out-performs OLS
and Elastic Net on next-month return prediction \emph{only} in the
AI/mega-cap regime, with $R^2_{\text{OOS}}=+1.29\%$ versus $+0.07\%$.
In the high-volatility regime XGBoost \emph{underperforms} linear
specifications, suggesting a covariate-shift problem during the
COVID-era training distribution.  The 2025 calendar year alone
delivers $R^2_{\text{OOS}}=+3.55\%$ for XGBoost on next-month return,
the largest single-year out-of-sample $R^2$ in the panel.  On
realized-volatility prediction, models are essentially equivalent and
$R^2_{\text{OOS}}$ exceeds $0.4$ in the AI/mega-cap regime.

\paragraph{3.~The dominant feature differs across regimes.}
Permutation feature importance reveals that the top predictor of
next-month returns differs across all three regimes:
open-interest-weighted vega in the low-volatility baseline, raw
post-filter contract count in the high-volatility regime, and OTM call
implied volatility in the AI/mega-cap regime.  None of the canonical
hand-engineered signals (smirk, IV spread, risk-neutral skewness) make
the top five in any regime once the model has access to the underlying
IV-surface and Greek primitives.  The fixed linear combinations that
define the canonical signals throw away information that nonlinear
models exploit, and which combinations matter depends on the regime.

\paragraph{4.~Crash forecasting works best when systematic risk is low.}
On a 5-day-forward firm-level crash flag, classification AUC is highest
in the low-volatility regime (XGBoost $0.706$, logit $0.670$), lowest
in the high-volatility regime (XGBoost $0.561$, logit $0.531$), and
intermediate in the AI/mega-cap regime ($0.665$ and $0.652$).  The IV
surface contains useful firm-level crash information when the
idiosyncratic component dominates and is largely uninformative when
macro shocks dominate.  XGBoost beats logit in every regime by
$0.01$--$0.04$ AUC.

\paragraph{Implications.}  Taken together, the results sharpen the
empirical claim that option-implied signals predict equity returns and
crash risk.  The canonical predictability is real but
\emph{regime-conditional}; the AI/mega-cap regime since 2023 looks
qualitatively different from the pre-COVID environment that established
the literature, and re-estimation on post-2022 data is essential
before importing the canonical effects into trading or risk-management
applications.  Machine-learning models that are trained jointly on the
underlying IV-surface and Greek primitives---rather than on the
hand-engineered linear combinations---recover predictability that the
canonical signals miss in the new regime.  Future work should
endogenize the regime classification, extend the ML benchmark to
deep architectures, incorporate signed retail flow, and test the
robustness of the regime-dependent feature rankings to alternative
target definitions.

The data, code, and intermediate parquet panels are available from the
author on request.  Replication is straightforward: the four-stage
pipeline (preprocessing, descriptives, panel regressions, ML benchmark)
runs end-to-end in approximately six hours of wall time on a 10-core
desktop with 16~GB of RAM, dominated by the panel construction step.

% --- Bibliography ----------------------------------------------------
\newpage
\bibliography{references}

% --- Appendix --------------------------------------------------------
\newpage
\appendix
% =====================================================================
\section{Variable definitions}
\label{app:variables}
% =====================================================================

This appendix consolidates the definitions of all features and outcomes.
All variables are at the firm-day level after the filters in
Section~\ref{sec:data} have been applied.  Index $i$ denotes the
underlying and $t$ denotes the trading day.  Sums and means within a
firm-day run over surviving option contracts on that firm-day; $K$ is
strike, $S$ is the underlying close, $\tau$ is days-to-expiry,
$\sigma$ is the model-supplied implied volatility, $V$ is daily
contract volume, OI is open interest, and $\Delta,\Gamma,\nu,\Theta$
are the standard Greeks.  Moneyness is $m\equiv K/S$.

\paragraph{IV-surface signals.}
\begin{itemize}
\item $\sigma^{\text{atm}}_{i,t}$: cross-contract mean of $\sigma$ over
contracts with $m\in[0.95,\,1.05]$.
\item $\sigma^{\text{atm,call}}_{i,t}$, $\sigma^{\text{atm,put}}_{i,t}$:
ATM mean restricted to calls or puts respectively.
\item $\sigma^{\text{otm,put}}_{i,t}$: cross-contract mean of $\sigma$
over puts with $m\in[0.80,\,0.95]$.
\item $\sigma^{\text{otm,call}}_{i,t}$: cross-contract mean of $\sigma$
over calls with $m\in[1.05,\,1.20]$.
\item $\sigma^{30}_{i,t}$, $\sigma^{60}_{i,t}$: ATM mean restricted to
$\tau\in[15,45]$ and $\tau\in[46,75]$ respectively.
\item smirk = $\sigma^{\text{otm,put}} - \sigma^{\text{atm,call}}$.
\item iv\_spread = $\sigma^{\text{atm,call}} - \sigma^{\text{atm,put}}$.
\item ts\_slope\_30\_60 = $\sigma^{60} - \sigma^{30}$.
\item rns\_bkm = $\frac{1}{n}\sum_{k=1}^{n}(\sigma^{\text{otm}}_k -
\bar\sigma^{\text{otm}})^3 / \hat\sigma^3$, where the index $k$ runs
over OTM (put or call) IVs of firm $i$ on date $t$ with $n\geq 5$.
\end{itemize}

\paragraph{Trading-activity signals.}
\begin{itemize}
\item pc\_volume\_ratio$_{i,t}$ = (sum of put volume) / (sum of call
volume).
\item pc\_oi\_ratio$_{i,t}$ = (sum of put OI) / (sum of call OI).
\item $n_{\text{contracts},i,t}$ = post-filter count of contracts.
\end{itemize}

\paragraph{Open-interest-weighted Greeks.}  For $g\in\{\Delta,\Gamma,
\nu,\Theta\}$:
\[
g^{\text{oi-w}}_{i,t} = \frac{\sum_k \text{OI}_k\,g_k}{\sum_k \text{OI}_k},
\]
falling back to an equal-weighted mean if the OI sum is zero.

\paragraph{Liquidity controls.}
\begin{itemize}
\item rel\_spread = $\frac{\text{ask}-\text{bid}}{\tfrac12(\text{ask}+\text{bid})}$,
averaged across contracts.
\item dollar\_spread = $\text{ask}-\text{bid}$, averaged.
\item frac\_zero\_volume = fraction of contracts with $V_k=0$.
\end{itemize}

\paragraph{Outcomes.}
\begin{itemize}
\item $r_{i,t+h}=\log(S_{i,t+h}/S_{i,t})$ for $h\in\{1,5,21,63\}$.
\item $\text{RV}^{(21)}_{i,t+21}$ = $\sqrt{252}$ times the rolling
21-day standard deviation of daily log returns over $[t+1,t+21]$.
\item $\text{crash}^{(5)}_{i,t} = \mathbb{1}[\,r^{(5)}_{i,t+5} <
-3.09\,\hat\sigma^{(252)}_{i,t}\sqrt5\,]$ where $\hat\sigma^{(252)}$
is the trailing-252-day daily-return standard deviation.
\item $\text{crash}^{(21)}_{i,t} = \mathbb{1}[\,r^{(21)}_{i,t+21} <
-3.09\,\hat\sigma^{(252)}_{i,t}\sqrt{21}\,]$.
\end{itemize}

\end{document}